\documentclass{article}

\usepackage{arxiv}

\usepackage[utf8]{inputenc} 
\usepackage[T1]{fontenc}    
\usepackage{hyperref}       
\usepackage{url}            
\usepackage{booktabs}       
\usepackage{amsfonts}       
\usepackage{nicefrac}       
\usepackage{microtype}      
\usepackage{lipsum}
\usepackage{natbib}
\usepackage[dvipsnames]{xcolor}

\usepackage{graphicx}

\title{Early Screening of SARS-CoV-2 by Intelligent Analysis of X-Ray Images}

\author{
    D. Gil, K. Díaz-Chito, C. Sánchez, A. Hernández-Sabaté \\
  Computer Vision Center - Dept. of Computer Science\\
  Universitat Aut\`{o}noma de Barcelona\\
  Cerdanyola del Vallès, 08193 \\
  \texttt{{debora}@cvc.uab.cat} \\
}

\begin{document}
\maketitle

\begin{abstract}

Future SARS-CoV-2 virus outbreak COVID-XX might possibly occur
during the next years. However the pathology in humans is so
recent that many clinical aspects, like early detection of
complications, side effects after recovery or early screening, are
currently unknown. In spite of the number of cases of COVID-19,
its rapid spread putting many sanitary systems in the edge of
collapse has hindered proper collection and analysis of the data
related to COVID-19 clinical aspects.

We describe an interdisciplinary initiative that integrates
clinical research, with image diagnostics and the use of new
technologies such as artificial intelligence and radiomics with
the aim of clarifying some of SARS-CoV-2 open questions. The whole
initiative addresses 3 main points: 1) collection of standardize
data including images, clinical data and analytics; 2) COVID-19
screening for its early diagnosis at primary care centers; 3)
define radiomic signatures of COVID-19 evolution and associated
pathologies for the early treatment of complications.

In particular, in this paper we present a general overview of the
project, the experimental design and first results of X-ray
COVID-19 detection using a classic approach based on HoG and
feature selection. Our experiments include a comparison to some
recent methods for COVID-19 screening in X-Ray and an exploratory
analysis of the feasibility of X-Ray COVID-19 screening. Results
show that classic approaches can outperform deep-learning methods
in this experimental setting, indicate the feasibility of early
COVID-19 screening and that non-COVID infiltration is the group of
patients most similar to COVID-19 in terms of radiological
description of X-ray. Therefore, an efficient COVID-19 screening
should be complemented with other clinical data to better
discriminate these cases.

\end{abstract}


\section{Introduction}

In the last months, coronavirus SARS-COV-2, which causes COVID-19,
has widely spread all over almost all the countries of the world.
Due to its high contagiousness, at the present time (19 May
2020\footnote{https://www.who.int/emergencies/diseases/novel-coronavirus-2019}),
the World Health Organization (WHO) confirms a total of 4,731,458
confirmed cases and 316,169 of total deaths all over the world in
less than five months, causing a collapse in several health
systems. Although the course of the disease is often mild, in a
considerable number of cases may lead to a severe pneumonia, among
other complications, that can rapidly get worse and require
intensive care. Therefore and since there is no effective
treatment for COVID-19 yet, early screening of the disease is
highly recommended for the identification and treatment high-risk
patients, thus, preventing a severe progression of the disease
\cite{sun2020lower}.

There is evidence that Chest CT has a sensitivity for diagnosis of
COVID-19
\cite{li2020artificial,ai2020correlation,wong2020frequency}. In
particular, patients with confirmed COVID-19 pneumonia have
typical imaging features that can be helpful in early screening
and in evaluation of the severity and extent of disease
\cite{zhao2020relation}. However, CT is very expensive and
difficult to make massively to the population. In contrast, X-ray
is a low cost modality based on the same technology as CT which is
available at primary care centers. Besides, X-ray allows an
affordable rapid triaging that is already being used at hospitals
to confirm the disease and monitor patients' recovery since it is
a highly available and affordable technique.

The goal of this study is to early diagnose and follow-up patients
with COVID-19 from an intelligent analysis of X-ray images. Thus,
in this paper we present our general approach for early screening
of COVID-19 using X-ray and the first results obtained using a
classical approach on a dataset extracted from several public
repositories. We describe the sampling strategy for defining a
training and testing sets and present the results obtained for
Histogram of Oriented Gradients (HoG) descriptor
\cite{dalal2005histograms} and a reduction of dimensionality.

Our experiments include the assessment of the most suitable method
for dimensionality reduction and HoG parameters, a comparison to
state-of-art methods and an exploratory analysis of the capability
for COVID-19 early detection. Results show that our approach can
outperform deep-learning methods, indicate the feasibility of
early COVID-19 screening and identify non-covid infiltration as
the group of patients most likely to be radiologically confused as
COVID-19. The latter suggests adding clinical variables in order
to increase the efficiency of a COVID-19 screening based on X-ray
analysis.

The remainder of the paper is structured as follows. Section
\ref{sc:SoA} summarizes the state-of-the-art related with the
paper. Section \ref{sc:projectOverview} overviews how artificial
intelligence can contribute to the fight against the pandemic.
Section \ref{sc:EarlyScreening} details the methodology used to
early detect COVID-19 in X-Ray images. Section
\ref{sc:experiments} is devoted to explain the experimental
setting and show the results assessing our method and, finally,
discussion and conclusions are provided in section
\ref{sc:discussion}.

\section{State of the art}\label{sc:SoA}

Early detection of COVID-19 from X-rays has risen great interest
within the artificial intelligence community. Although results
seem encouraging, we consider that there are several issues
related to, both, the available data and the experimental design
that should be taken into account for fair interpretation of
results.

\begin{figure}
\begin{tabular}{ccc}
     \centering
\includegraphics[width=.3\linewidth]{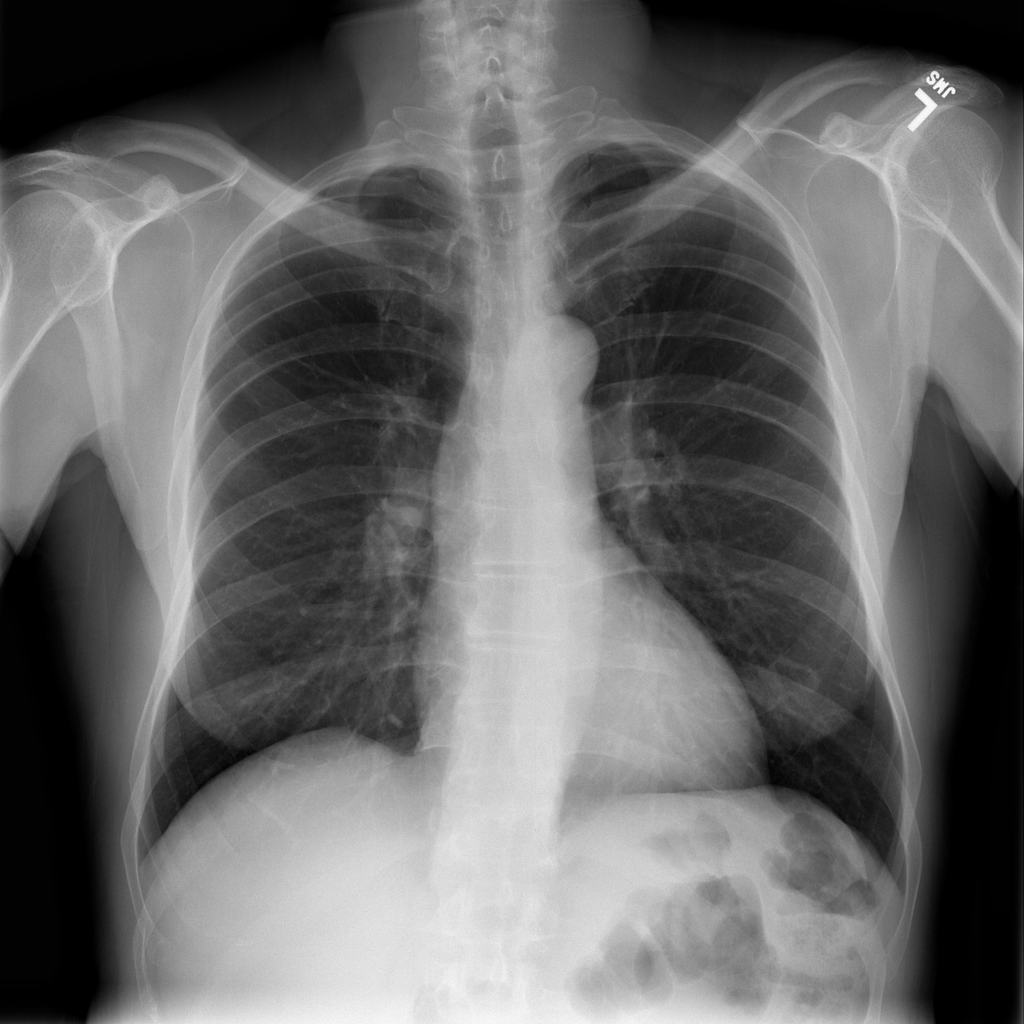}
&
\includegraphics[width=.3\linewidth]{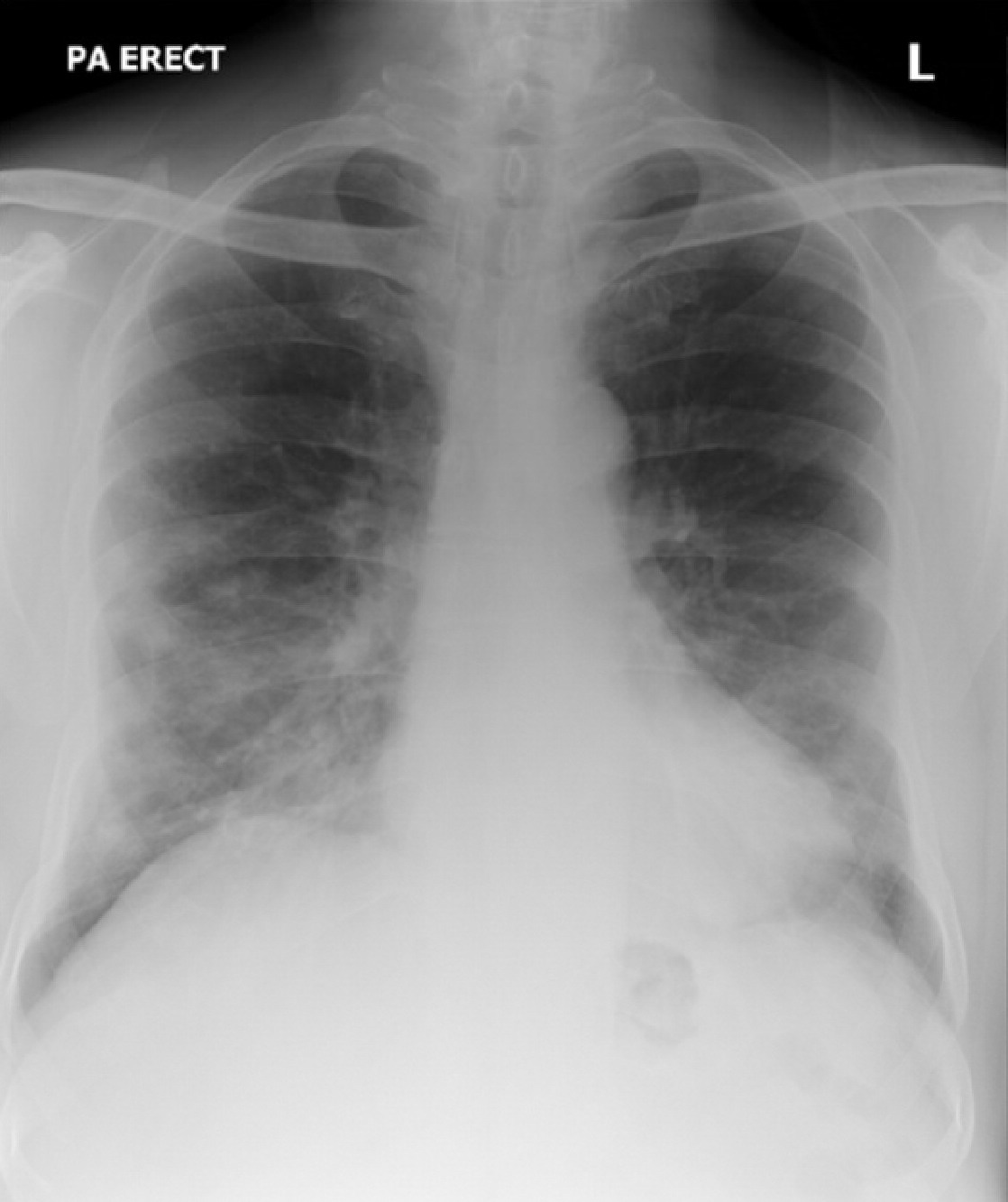}
&
\includegraphics[width=.3\linewidth]{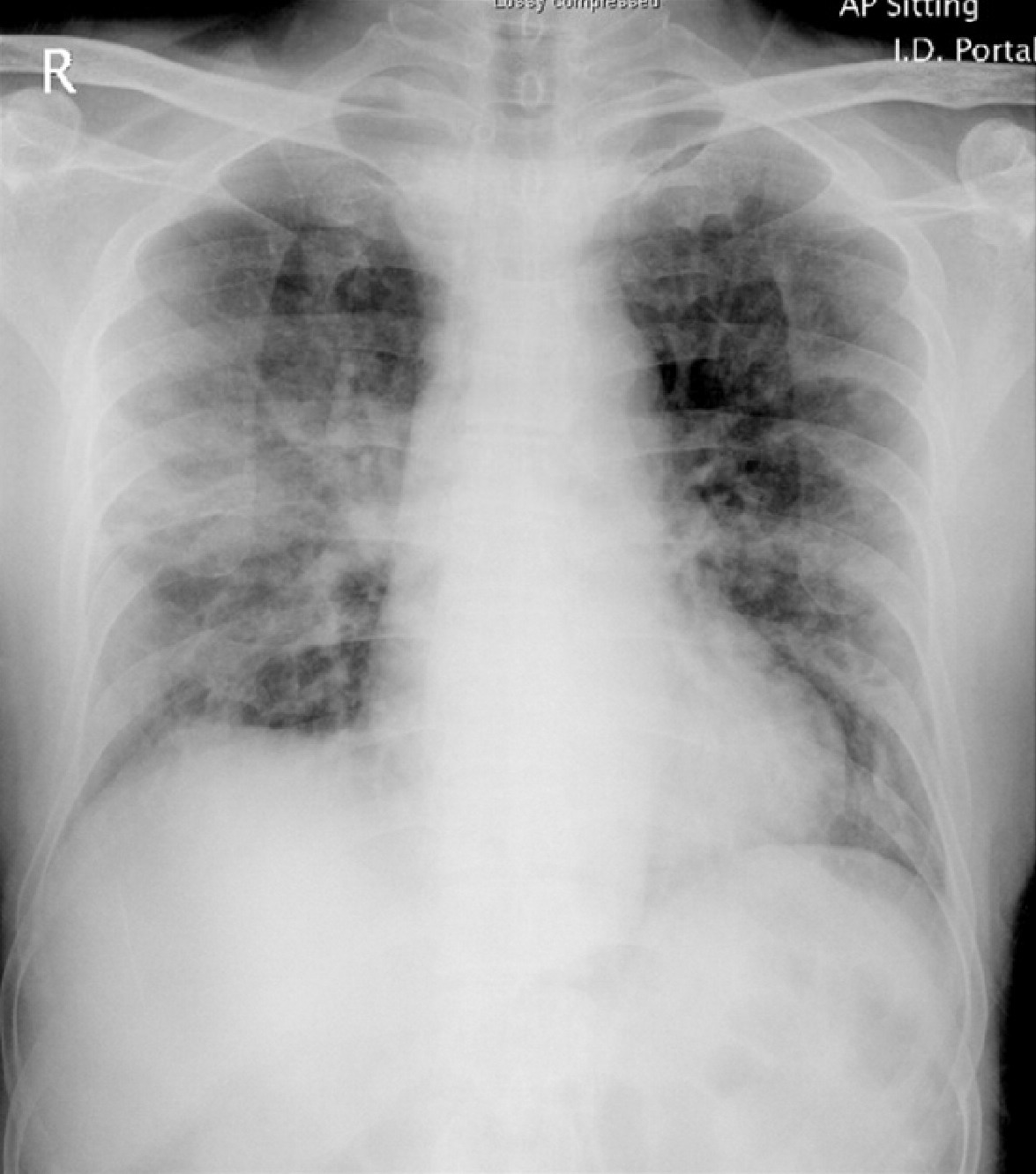}\\
(a) & (b) & (c)
\end{tabular} \caption{Comparison between Normal and COVID
X-ray images: normal case, (a) COVID at early stage, (b), COVID at
advanced stage, (c).} \label{fig:NormalvsCOVID}
\end{figure}

The main radiological feature of COVID-19 (fig.
\ref{fig:NormalvsCOVID} (b), (c)) is the development of pneumonia.
Radiologically, pneumonia produces light (white and gray) areas
(related to tissue inflammation) inside lungs, which usually show
dark due to the low density of their tissue (see fig.
\ref{fig:NormalvsCOVID} (a)). Aside COVID-19, there are several
pathologies (like edema or non-COVID pneumonia) having similar
radiological description. Unlike other pathologies with similar
radiological description, COVID-19 pneumonia has a rapid
progression (compare images in fig. \ref{fig:NormalvsCOVID} (b)
and (c)) prone to collapse lungs. At this point, they become
completely white in X-ray (see fig. \ref{fig:NormalvsCOVID} (c)),
loose their functionality and the patient requires immediate
intensive care assistance. Therefore, a computational system
detecting COVID-19 at advanced stages, although probably very
accurate, lacks of any clinical value. This implies that data for
training and testing methods should be carefully chosen and
include, both, COVID-19 early stages and all radiological types
similar to COVID-19 associated to other pathologies. Under these
considerations, we do a critical review of existing databases and
methods.

The most extended public database with COVID+ cases is the one
explained in \cite{cohen2020covid}. At this moment, it contains
360 frontal view X-rays and CT from 199 different patients with
different respiratory pathologies. Images have been extracted from
online publications, website, or directly from PDFs, maintaining
the quality of the images.  The database also contains metadata
that includes, for some cases, how many days after hospitalization
of the patients images were acquired. Although these offset
variable could be considered as an approximation to COVID-19
stage, more clinical data is needed to fully assess COVID-19
stage. The dataset in \cite{andrewmvd} contains 20 frontal and
lateral X-ray images and CT snapshots of patients diagnosed with
COVID-19. All images were extracted from publicly available
articles and knowledge bases. The one published in
\cite{paulmooney} contains 5863 images of X-ray images with and
without pneumonia (no COVID-19). The data base of the Italian
society of medical and interventional radiology (SIRM)
\footnote{http://www.sirm.org/category/senza-categoria/covid-19/}
deserves special attention because it contains clinical data,
including symptoms, patient data, diagnosis and radiological
description from 219 COVID-19 positive images, 1341 normal images
and 1345 viral pneumonia images. Although highly valuable from a
clinical point of view, a main inconvenience for its use in an
artificial intelligence system is that radiographic images have
varying resolution and clinical data has not been collected in a
systematic manner. Data are the clinical annotations of the
physician or radiologist who attended the patient. Thus, it does
not always contain the same information and some language parser
should be used to identify the different vocabulary used to
describe the same pathology.

A main concern about the above data sources is that they are
neither standardized nor record patient's clinical data such as
whether the disease is at an early or advance state. The latter
information is mandatory for early diagnosis and follow-up of the
disease. As well, some of them (such as \cite{chowdhury2020can})
develop their database from other databases (such as
\cite{cohen2020covid}) so that merging them to obtain more images
does not guarantee non-overlaps between them.

Regarding the methodologies used so far, all of them pose COVID-19
detection as a classification problem. Most use transfer learning
to explore the discriminative capability of existing Convolutional
Neural Networks \cite{apostolopoulos2020covid, narin2020automatic,
hemdan2020covidx, sethy2020detection, zhang2020covid} already
trained with non-clinical databases. Some works replace the deep
classifier with a classic one such as SVM
\cite{sethy2020detection, castiglioni2020artificial} or combine
the deep classifier with an anomaly detector
\cite{zhang2020covid}. The only work presenting an architecture
specific to detect COVID-19 is the one described in
\cite{wang2020covid}. However, they use again transfer learning
and pre-train the network using ImageNet. Concerning the
definition and training of the classification problem, most
methods
\cite{apostolopoulos2020covid,narin2020automatic,hemdan2020covidx,sethy2020detection,
zhang2020covid,castiglioni2020artificial,wang2020covid} define a
two class problem (COVID/noCOVID) with the noCOVID class including
normal and non-COVID pneumonia for
\cite{apostolopoulos2020covid,sethy2020detection,
zhang2020covid,castiglioni2020artificial,zhang2020covid} and only
normal cases for \cite{narin2020automatic,hemdan2020covidx}. The
only method training a multiclass problem including normal,
nonCOVID bacterial and nonCOVID viral pneumonia and COVID is the
own design network described in \cite{wang2020covid}. None of the
existing works includes cases with lung infiltration, which
radiologically look very similar to a COVID since they are typical
in viral pneumonia but can also be originated by other
pathologies.

Table \ref{tb: Methodology used} details the methodology used in
the papers found in the bibliography. In particular, we detail the
dataset they use (2nd column), CNN/feature extraction they use,
process of transfer learning, if they balance the data or not and
the classes considered.

Although results seem promising in most works, there are several
issues that should be pointed out. First, the unavailability of a
large amount of data is a problem to ensure the reliability of
transfer learning, since the images with which networks have been
pre-trained can have very different characteristics than we are
looking for. Second, the groups considered for the classification
problem might not be the most appropriate for COVID-19 early
detection. Binary COVID/Normal approaches like
\cite{narin2020automatic,hemdan2020covidx} can not assess
discrimination between Pneumonia caused by other sources (like
bacteria or non-COVID virus), which it is crucial for a reliable
screening program. The remaining approaches lack of some
radiological descriptions (like infiltration) that are present in
viral pneumonia but also in other pathologies and, thus, their
capability for discriminating among these pathologies remains
unknown. Finally, none of the approaches reports the stage of the
COVID-19 cases used. Given that advanced stages are visually
easily identified (fig. \ref{fig:NormalvsCOVID} (c)), these facts
can introduce an optimistic bias in the reported accuracy
measures.

\begin{table}[htp!]
\begin{center}
\begin{tabular}{ | p{0.10\textwidth} | p{0.2\textwidth} |  p{0.2\textwidth} |  p{0.15\textwidth}|  p{0.12\textwidth} | p{0.12\textwidth} | }
\hline
 \textbf{Article} & \textbf{Dataset} & \textbf{Feature extraction}& \textbf{Transfer learning} & \textbf{Data balance}&\textbf{Classes}\\
 \hline
\cite{apostolopoulos2020covid} & Dataset 1: 224 COVID-19 \cite{cohen2020covid} +
700 common bacterial pneumonia + 504 normal\cite{andrewmvd} & VGG19, MobileNet v2,
Inception, Xception, InceptionResNet v2 & last layer&No& CV/ PNE/ N\\
& Dataset 2: 224 COVID + 714 (400 bacterial +314 viral) pneumonia + 504 normal &  &&& \\
  \hline
 \cite{narin2020automatic} & 50 COVID \cite{cohen2020covid} + 50 normal \cite{paulmooney}
 & Inception v3, ResNet50, InceptionResNet v2 & 1 global avg pooling 2D + 2 FC layers&Not reported& CV/ N\\
  \hline
  \cite{hemdan2020covidx} &25 COVID \cite{cohen2020covid} + 25 normal \cite{paulmooney}
  & DenseNet201, VGG19, Inception v3, Xception, MobileNet v2, ResNet v2, InceptionResNet v2
  & tuning deep learning classifier &Not reported&CV/ N\\
  \hline
  \cite{sethy2020detection} &collected from GitHub,
Kaggle and Open-I repository with unknown number of images
& AlexNet, VGG16, VGG19, GoogleNet, ResNet18, ResNet50, ResNet101, Inception v3,
InceptionResNet v2, DenseNet201, Xception& 1 FC layer + SVM & Not reported & CV/ PNE\\
 \hline
\cite{zhang2020covid}& 100 x-ray images across 70 patients with
COVID-19 \cite{cohen2020covid} + 1431 x-ray images across 1008
patients with pneumonia no COVID \cite{wang2017chestx} & Backbone
network: Residual Neural Network +
 trade off between SVM classifier and anomaly detector & not reported &Augmentation& CV/ PNE\\
\hline \cite{wang2020covid} & 5941 posteroanterior chest
radiography images across 2839 patient cases (45 COVID + 1203
normal + 931 bacterial pneumonia + 660 non-COVID viral
pneumonia)\cite{cohen2020covid}&
COVID\_Net pre-trained with ImageNet  & Not reported & No  & CV/ PNE/ N\\
\hline \cite{castiglioni2020artificial} &Training: 250 COVID + 250
no COVID, Testing: 74 COVID+ 36 no COVID
&ResNet50& last layers, but not reported which ones & Not reported & CV/ PNE\\
\hline
\end{tabular}
\caption{Summary of the methodologies used in the state of the
art. Different classes are identified as: CV (COVID-19), N (Normal), PNE (non COVID-19 Pneumonia). } \label{tb: Methodology used}
\end{center}
\end{table}

Another issue introducing a bias in results, concerns the
experimental design used to validate and train methods. A main
concern is that there is no standardized validation protocol,
which hinders fair comparison across methods. The experimental
design for validation should include a description of the data
partition used for training and test, the quality scores for
assessment of performance and the statistical analysis carried
out.

While some of them assess performance using a k-fold
cross-validation \cite{apostolopoulos2020covid,
castiglioni2020artificial}, others split data into a single
training and test subsets \cite{narin2020automatic,zhang2020covid}
and some directly do not explain the procedure used
\cite{hemdan2020covidx, sethy2020detection, wang2020covid}. The
only partition that allows statistical analysis of quality scores
is a k-fold design.

Concerning quality measures, the most informative scores for
clinical assessment of COVID-19 screening are sensitivity (also
called recall) and precision (or Positive Predictive Value - PPV).
The first one quantifies the percentage of COVID-19 cases that the
system detects correctly. The second one quantifies the percentage
of cases wrongly detected by the system as COVID-19 from all
COVID-19 cases detected. This is relevant if diagnosed cases
require further tests in order to confirm the pathology. The
remaining scores reported in the literature are not so
informative, given that we have a highly unbalanced problem with
COVID-19 the minority class. A main concern is that none of the
existing studies performs any assessment of the capability for
early COVID-19 detection.

Finally, all works (with the exception of
\cite{castiglioni2020artificial}) lack of any statistical
analysis. Comparison across the different networks is simply done
using average scores, which have not any statistical significance
and, thus, it is not guaranteed reproducibility of results in new
cases.

Table \ref{tb: experimental design} details the performance
evaluation of the methods reviewed in the state of the art. For
the performance procedure we also show the data partition in case
the authors detail it.

\begin{table}[htp!]
\begin{center}
\begin{tabular}{ | p{0.1\textwidth} | l | p{0.2\textwidth}  | p{0.09\textwidth} | p{0.09\textwidth}  | p{0.09\textwidth}  | p{0.09\textwidth} |}
\hline
 \textbf{Article} & \textbf{Network} & \textbf{Performance evaluation procedure} & \textbf{Accuracy}& \textbf{Sensitivity (Recall)} & \textbf{Specificity}&\textbf{Precision}\\
 \hline
 \cite{apostolopoulos2020covid} & VGG19 Dataset 1& 10-fold cross-validation & {\bf 98.75} & 92.85 & {\bf 98.75} & {\bf 93.27}\\
 & MobileNet v2 Dataset 1& &97.40 & {\bf 99.10} & 97.09 & 86.38\\
 & MobileNet v2 Dataset 2 & &96.78 & 98.66 & 96.46 & NR\\
  \hline
 \cite{narin2020automatic} & Inception v3 & 5-fold cross-validation  & 97 & 94 & {\bf 100} & {\bf 100}\\
 & ResNet50 & &{\bf 98} & {\bf 96} & {\bf 100}& {\bf 100}\\
 & InceptionResNet v2 & &87 & 84 & 90 & 91\\
  \hline
  \cite{hemdan2020covidx} &VGG19 & Train: validation: test  40\%:40\%:20\% & 90 & {\bf 100} & 80 & 83\\
  & DenseNet201 & &90 & {\bf 100} & 80 & 83\\
   & ResNet v2 & &70 & 40& {\bf 100} & {\bf 100}\\
   & InceptionResNet v2 && 80 & 60 & {\bf 100} & {\bf 100}\\
   & Xception & &80 & 60 & {\bf 100} & {\bf 100}\\
   & MobileNet v2 & &60 & 20 & {\bf 100} & {\bf 100}\\
  \hline
  \cite{sethy2020detection} & ResNet50 & Train: validation: test 60\%:20\%:20\%, 100 simulations & 95.38 & 97.29 & 93.47 & NR\\
 \hline
\cite{zhang2020covid}&Backbone & 2-fold cross-validation& NR & 96 & 70.65 & NR\\
\hline
\cite{wang2020covid} & COVID\_Net & NR & 83.5 &100 &NR&80\\
\hline
\cite{castiglioni2020artificial}&ResNet50&10-fold cross-validation&80&79.72 CI = (72,86)& 80.55 CI =(73,87)&89.39 CI = (82,94)\\
\hline
\end{tabular}
\caption{Summary of the experimental design used in the state of
the art. NR are not reported results.} \label{tb: experimental design}
\end{center}
\end{table}

\section{Project Overview}\label{sc:projectOverview}

This project \footnote{http://iam.cvc.uab.es/portfolio/covair/} is
a collaborative initiative between the Computer Vision Center, and
the Research Unit Support of IDIAP Jordi Gol for early diagnosis
and follow-up of COVID-19 patients from X-ray imaging analysis. In
particular, our initiative has three main objectives:

\begin{enumerate}

\item {\it Compile a Standardized Database of COVID-19}. A
COVID-19 standardized x-ray database will be collected that
includes COVID-19 and non-COVID-19 cases along with the clinical
and population data required for computational analysis.

\item {\it COVID-19 Early Diagnosis}. The goal is to classify
x-ray images to discriminate COVID-19 pneumonia from other types
of pneumonia. This way, we could provide a screening tool for
early and rapid diagnosis at the primary health care level.

\item {\it COVID-19 Monitoring}. Identify at very early stages
normal/non-normal X-ray images with infiltrations and find visual
progression patterns that may be characteristic of COVID-19 for
predicting possible complications requiring hospitalization.

\end{enumerate}

\subsection{Data collection}

Patients who have had a chest x-ray in a primary care facility and
who have suspected symptoms of COVID-19 will be collected. The
following groups of patients will be recruited:
\begin{enumerate}
    \item Patients with COVID and pneumonia (COVID-19 pneumonia),
    \item Patients with COVID-19 and no pneumonia (control group),
    \item Patients without COVID-19 and pneumonia (non-COVID pneumonia),
    \item Patients suspected of having COVID-19 (unconfirmed) and pneumonia,
    \item Patients with respiratory symptoms but no pneumonia (second control group),
    \item Patients recovered from COVID pneumonia,
    \item Patients recovered from other types of pneumonia (after non-COVID-19 pneumonia).
\end{enumerate}

Data will be collected from prospective and retrospective cases
from Catalan primary care centres.

Apart from the imaging data, the following information will also
be collected for statistical analysis of population factors: date
of each radiograph, date of initial and final diagnosis, gender,
age and centre where the images were acquired, and clinical data
related to the electronic medical records of each patient. The
latter will include the complications that each patient has had in
order to develop objective 3 of this project.

\subsection{COVID-19 Early Diagnosis}

For the diagnostic imaging system, a radiomic artificial
intelligence system shall be developed to simultaneously analyze
X-ray studies and clinical data. The image analysis system shall
be defined to distinguish the different groups and the following
technical options shall be considered:

\begin{enumerate}
    \item Classical classifier \cite{duda2012pattern}
based on visual characteristics defined by formulae and filter
banks. Although these techniques have a lower accuracy than
deep-learning approaches, models can be adjusted with few data.
For this system, a radiomic feature space will be defined
\cite{lambin2017radiomics} that includes local texture and shape
descriptors. Classic radiomic feature spaces such as pyradiomics
\cite{van2017computational}, as well as texture descriptors for
pattern analysis such as HoG \cite{dalal2005histograms} or LBP
\cite{ojala1996comparative} will be considered. To compensate for
the low number of samples (cases), a selection of the most
discriminating measurements will be made using linear (PCA, LDA,
mRMR) and non-linear (KDCV, KPCA) methods. On this space of low
dimensionality characteristics, classical classifiers such as SVM
will be applied.

\item Classifier based on deep-learning
\cite{schmidhuber2015deep}. The number of cases that can be
collected in this study discourages a trained model from scratch.
Transfer learning strategies will be adopted, but in this case the
networks will be trained with public databases of respiratory
pathologies (e.g. ChestXray-NIHCC) including some COVID-19 cases
(e.g. covid-chest-xray) to then test the features of the deeper
layers in the data collected in this study. For the network
architecture, we will consider both an auto-encoder trained with
non COVID-19 cases to detect COVID-19 cases as anomalies, and a
classification network on the different pathologies including
covid-chest-xray COVID cases.
\end{enumerate}

\subsection{COVID-19 Monitoring}

For the system of prediction and prevention of COVID
complications, a radiomic signature \cite{lambin2017radiomics}
will be defined that considers sequentially all the plates of the
same patient.  As for the diagnostic system, two approaches will
be considered:
\begin{enumerate}
    \item Classical radiomic signature \cite{lambin2017radiomics}.
The descriptors (both radiomic and texture and shape descriptors,
as well as deep descriptors) of the diagnostic system will be
considered for all sequentially X-rays. A dimensionality reduction
that suppresses correlated variables (mRMR) and a logistic
regression model with a penalty (such as LASSO or elastic net)
will be applied to define the signature most correlated with risk
of complication. Different models will be made for each time
sequence of x-ray images to determine the minimum number needed to
predict a negative evolution of COVID-19 patients.

\item Signature based on neural networks. Recurrent neural
networks \cite{schmidhuber2015deep} allow the analysis of data
sequences and have proven to be effective in a variety of domains
(natural language processing, handwriting recognition or genome
sequence analysis). As the number of cases that can be collected
in this study discourages a convolutional model trained from
scratch on the images themselves, a recurrent network will be
trained on the visual descriptors extracted for diagnosis. This
will allow the definition of simplified architectures omitting the
convolutional part and having as input the one-dimensional vector
defined by these descriptors. As in the previous point, networks
will be trained with both classical and deep descriptors and the
number of recurrent layers will be optimized to determine the
minimum number of x-ray images needed to predict a complication.
\end{enumerate}

For both objectives, clinical data will be incorporated to detect
factors correlated with COVID-19 and increase the discriminative
capacity of image analysis. The techniques developed within the
Up4Health project led by CVC\footnote{http://iam.cvc.uab.es/} will
also be applied to decrease the impact of variability in input
data associated with acquisition protocols and increase the
reproducibility of the diagnostic system. With the aim of
generating efficient and adaptable models, an incremental learning
architecture will be developed that is capable of evolving the
learning model for future mutations of the COVID-19. This
incremental approach will be based on dual learning \cite{Kate19},
which allows for real time information processing (e-learning).

\subsection{Statistical Analysis}

To validate which AI strategy gives the best system of diagnosis
and prediction of complications, sensitivity, specificity and
precision in the detection of COVID-19 and non-COVID-19 cases will
be calculated. For each of these quality indicators, an analysis
of variance statistical test will be performed to detect
differences between the different strategies. In addition,
generalized regression models will be used to identify significant
clinical factors that correlate with the diagnosis.

\section{An Exploratory Approach to Early Detection of COVID-19}\label{sc:EarlyScreening}

From the point of view of machine learning, COVID-19 detection is
a small size unbalanced problem \cite{Fukunaga,Cohen18}, being the
target class COVID-19 the minority one. Such condition poses a
main challenge for accurate performance of machine learning
strategies, including deep learning methods. While this can be
mitigated by acquiring larger datasets to balance the ratio, being
COVID-19 a recent pathology, this is not possible at the moment.

Data augmentation \cite{Krizhevsky12} has become a standard
procedure to improve the training process. Data augmentation
schemes increase the number of training samples by simple
transformations (like translation, rotation, ip and scale) of the
original dataset images. However, the diversity that can be gained
from such modifications of the images is relatively small and
introduces correlations in training data. These artifacts are
prone to drop the reproducibility of machine learning methods,
especially in the case of clinical predictions \cite{Li16}.

An alternative to mitigate the curse of dimensionality, it is to
apply a dimensionality reduction and feature extraction method
\cite{Kate19}. Given the small size of the COVID-19 class in
public data bases, we have adopted the latter in a classical
approach using hand-crafted features and a SVM classifier
\cite{SVM} with the aim of exploring the feasibility of early
COVID-19 detection in X-ray.

In the next sections we explain how we have defined a balanced
data set from public data repositories and the proposed feature
space.

\subsection{DataSet Definition from Public
Repositories}\label{Sec:DataBase}

In this study we have used cases from 2 public repositories: the
kaggle database
\footnote{https://www.kaggle.com/bachrr/covid-chest-xray} and the
NIH Chest X-ray Dataset of 14 Common Thorax Disease Categories
\footnote{http://academictorrents.com/details/557481faacd824c83fbf57dcf7b6da9383b3235a}
\cite{Cohen2014,Lo2016}.

The repository \cite{cohen2020covid} also in kaggle is the most
widely used database for COVID-19 detection in X-ray. This
database contains front X-ray views from pneumonia caused by
several pathogens, including viral and bacterial and an Excel file
with metadata of each image. Images have been extracted from
online publications, web pages or directly from PDFs, trying to
maintain the quality of the images. The data base currently
contains 288 X-ray front views from 62 cases (patients) with
confirmed COVID-19. For some of the patients, several images taken
at different dates are available. The number of days since the
start of symptoms or hospitalization for each image is stored in
the field offset of the metadata file. This is very important to
track progression of the pathology.

The second database \cite{wang2017} comprises 112120 frontal-view
X-ray images of 30805 unique patients with 14 common thoracic
pathologies including Atelectasis, Consolidation, Infiltration,
Pneumonia, Edema, Emphysema, Fibrosis, Effusion, Pneumonia,
Pleural Thickening, Cardiomegaly, Nodule, Mass and Hernia. The
database also includes normal cases labelled as Non-Finding.

\begin{figure}
\begin{tabular}{cccc}
     \centering
\includegraphics[width=.25\linewidth]{Normal.png}
&
\includegraphics[width=.25\linewidth]{covid-day4.jpeg}
&
\includegraphics[width=.25\linewidth]{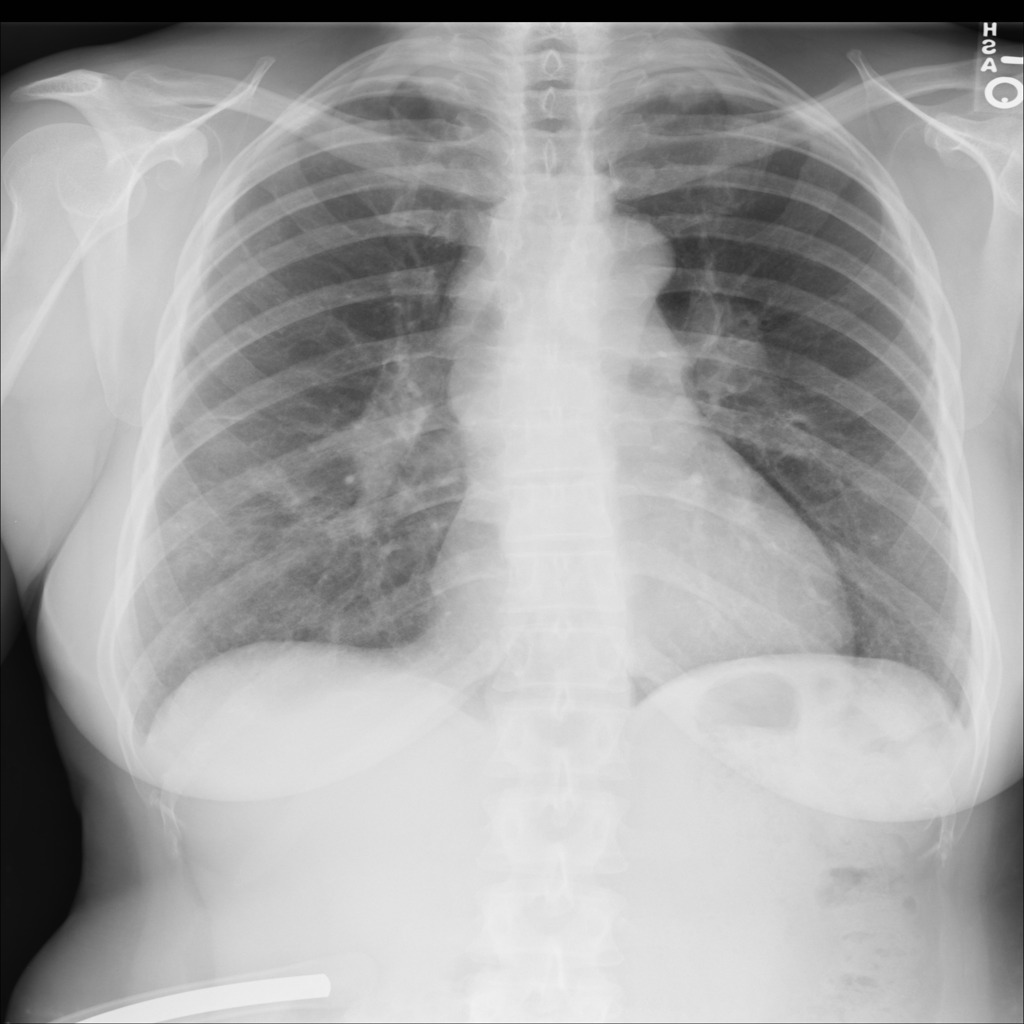}&
\includegraphics[width=.25\linewidth]{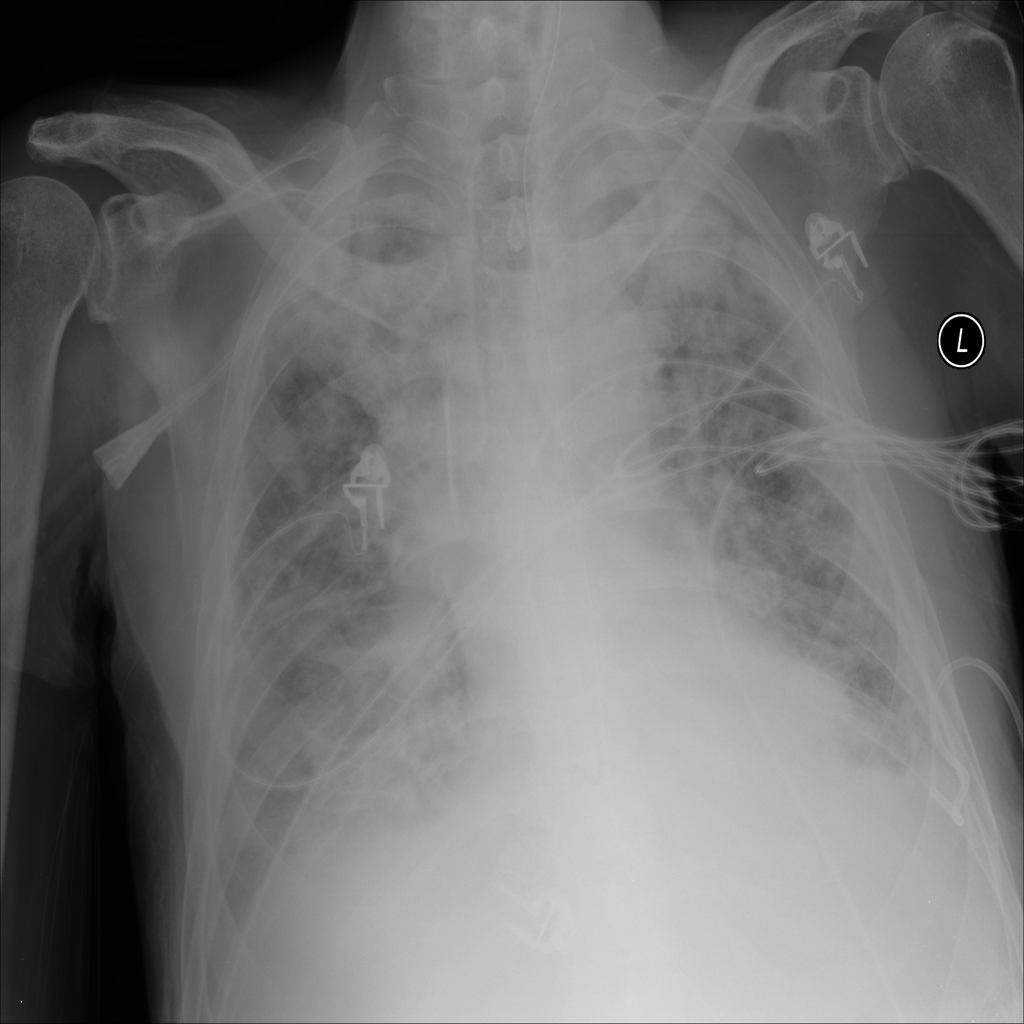}\\
(a) & (b) & (c) & (d)
\end{tabular} \caption{Examples of X-ray images for each class in the data base: normal case,
(a) COVID at advanced stage, (b), non-COVID infiltration, (c) and
non-COVID pneumonia, (d).} \label{fig:DataBaseClasses}
\end{figure}

To conduct this study, we have created our own dataset containing
all cases from kaggle classified in 2 categories (COVID-19 and
pneumonia) and a sub-sampling of the NIH Chest X-ray Dataset
Infiltration, Pneumonia and Non-Finding. These groups were
selected by their potential radiological similarity to COVID-19.
Such sub-sampling was randomly selected and of size to balance
classes. In total our data set contains 1152 images of COVID-19,
Pneumonia, Infiltration and Non-Finding in frontal-view, with 288
images for each class. From now on, these classes will be labelled
COVID-19, non-COVID-19 pneumonia, non-COVID-19 infiltration and
Normal, respectively. Images have been scaled to $400 \times 400$
pixels and normalized in the range $[0,1]$. Figure
\ref{fig:DataBaseClasses} shows an example of X-ray image for each
class.

\subsection{Feature Space Definition}

By its capability to describe, both, shape and texture with a low
number of parameters we have chosen the HoG descriptor. The
technique counts occurrences of gradient orientation in a
partition of the image into cells. Counting in each cell is given
by the histogram of gradient orientations which are concatenated
for all cells to define HoG descriptor.

The only critical parameter of HoG is the size of the cell, since
it determines the level of detail (e.g. scale) HoG is able to
codify. Also this parameter determines the dimension of the HoG
feature space, which equals to $(NRow*NCol)/(CellSze^2)*NBins$,
for $NRow$, $NCol$ the rows and columns of the images, $CellSze$
the size of the HoG cell and $NBins$ the number of bins of the
histogram of gradients.

Regarding the reduction of dimensionality, we have considered the
following methodologies well-suited in case of a number of samples
smaller than the dimensionality of the feature space: Principal
Component Analysis (PCA), Kernel Principal Component Analysis
(KPCA), Linear Discriminant Analysis (LDA) and Discriminant Common
Vector (DCV) \cite{Cevikalp05}.

\section{Experiments}\label{sc:experiments}

In this exploratory study, 4 experiments have been conducted to
assess the feasibility of COVID-19 screening using X-ray:

\begin{enumerate}

\item {\it Determine the optimal method for reduction of
dimensionality.} For this experiment HoG was applied with
$CellSze=4$. In order to ensure maximum inter class separability,
DCV was applied with a fraction of variance to form the
pseudo-null space equal to 0.8.

The quality of the reduction of dimensionality was visually
assessed by plotting the classes in the spaced defined by the 3
principal components of each method. Reduction methods were
trained using 60\% of the data. Visual assessment of the
distribution of the remaining test set in the reduced space was
enough to discard methods and select the best posed method for the
remaining experiments.

\item {\it Set the optimal scale for HoG descriptor.} In order to
set the most appropriate cell size, we have computed HoG with
$CellSze \in [4,8,16,32]$. The HoG space was reduced using the
method selected in the first experiment.

In order to select the optimal HoG scale, we have computed
precision (or PPV) and recall (also known as sensitivity). These
scores are commonly used in medical imaging applications since
they can measure the accuracy in pathology detection in unbalanced
settings. If we note $TP$ the number of true positives, $FP$, the
number of false positives, $TN$ the number of true positives and
$FN$ the number of false negatives, then precision and recall are
given by:
\begin{equation}
Precision=\frac{TP}{FP+TP} \mbox{      } Recall=\frac{TP}{TP+FN}
\end{equation}
Data was managed and analyzed with the software R, version 3.2.5.
A different generalized mixed linear mode of the effect of each
cell size was constructed for each quality. Models included the
fold as random effect. We calculated the 95\% CI of all scores and
p-values. A p-value < 0.05 was considered statistically
significant.

\item {\it Comparison to SoA.} We have compared our method to the
recent approaches based on deep learning summarized in Table
\ref{tb: Methodology used} with the scores reported in Table
\ref{tb: experimental design}. For the sake of a comparison as
fair as possible, we trained 3 different reductions of
dimensionality of our HoG with the cell size selected in the
second experiment with the classes used for each method in Table
\ref{tb: experimental design}: 1) COVID-19/Normal for
\cite{narin2020automatic,hemdan2020covidx}; 2)
COVID-19/non-COVID-19 Pneumonia for
\cite{castiglioni2020artificial,zhang2020covid} and 3)
COVID-19/non-COVID-19 Pneumonia/Normal for
\cite{apostolopoulos2020covid,wang2020covid}. For each class
configuration, we computed average scores for a 10-fold partition.

\item {\it Capability for COVID-19 early detection.} The
capability for early detection was tested by statistical analysis
of variance (anova) of COVID-19 detection with data grouped
according to the offset into 3 COVID-19 stages: early COVID-19
(offset <=3), mid COVID-19 (offset between 3 and 10) and late
COVID-19 (offset>10). The number of samples for each group was,
respectively, 18, 44 and 16. As before, a p-value < 0.05 was
considered statistically significant.

In order to compute COVID-19 detections in the whole data set, we
aggregated the evaluation of all k-fold models on their COVID-19
test sets. The boolean variable given by a correct detection was
the input for the analysis of variance.

\end{enumerate}

The code for all the experiments conducted in this paper is
available at https://github.com/IAM-CVC/CovAIR. This repository contains
Matlab code for the definition of the data base described in
\ref{Sec:DataBase}, as well as, one script for each of the
experiments. Methods require Matlab PRTools 5.0 toolbox.

\subsection{Results}

\begin{figure*}[!ht]
\begin{tabular}{cc}
     \centering
     \includegraphics[width=0.4\linewidth]{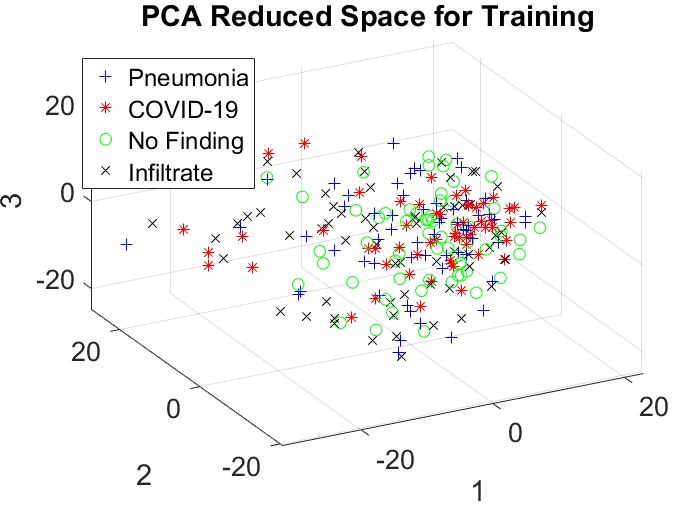}
     &
     \includegraphics[width=0.4\linewidth]{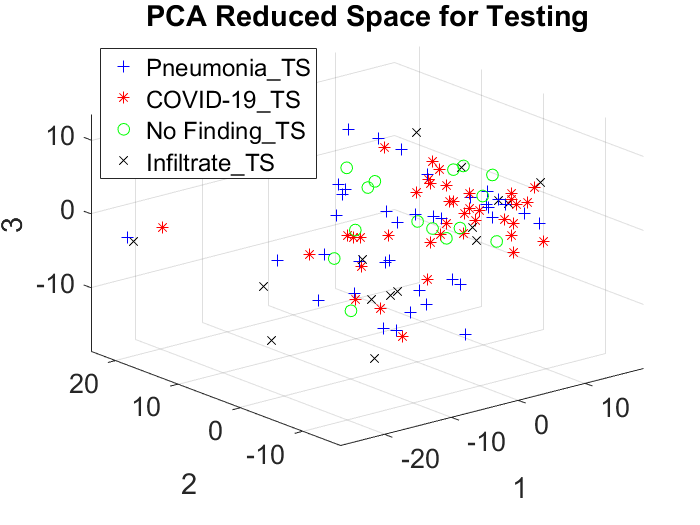}\\

     \includegraphics[width=0.4\linewidth]{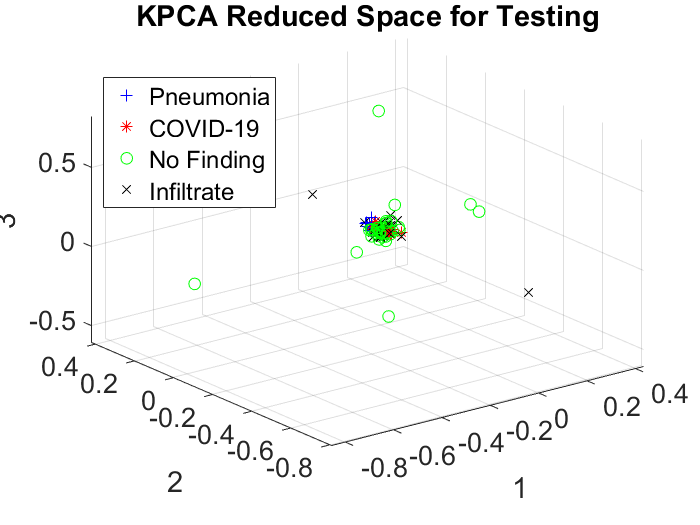}
     &
     \includegraphics[width=0.4\linewidth]{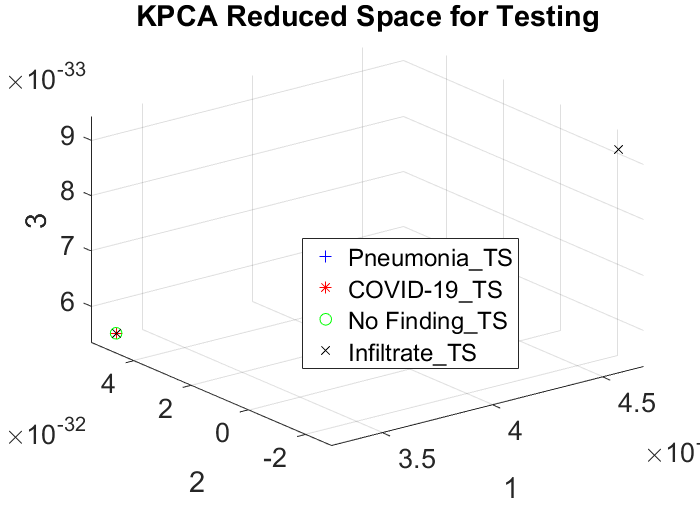}\\

     \includegraphics[width=0.4\linewidth]{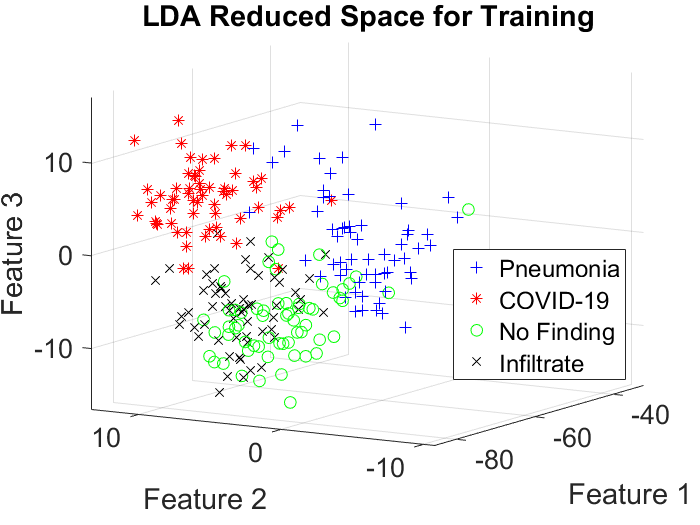}
     &
     \includegraphics[width=0.4\linewidth]{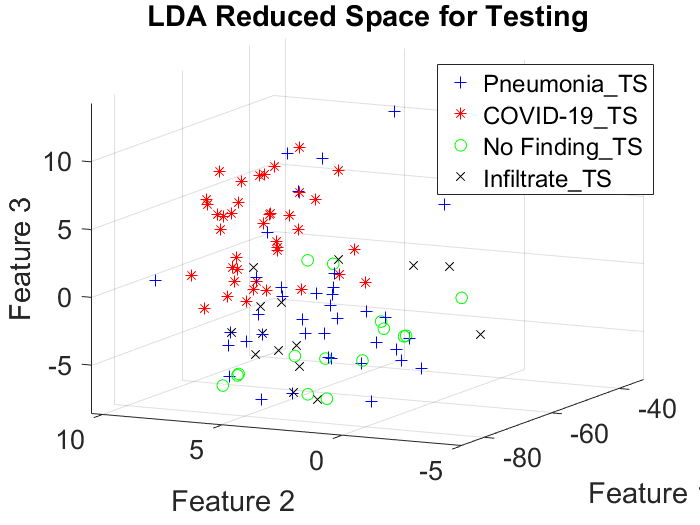}\\

     \includegraphics[width=0.4\linewidth]{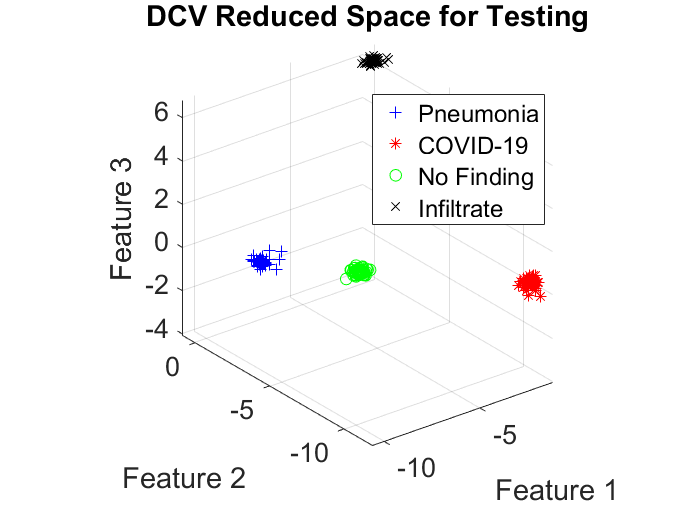}
     &
     \includegraphics[width=0.4\linewidth]{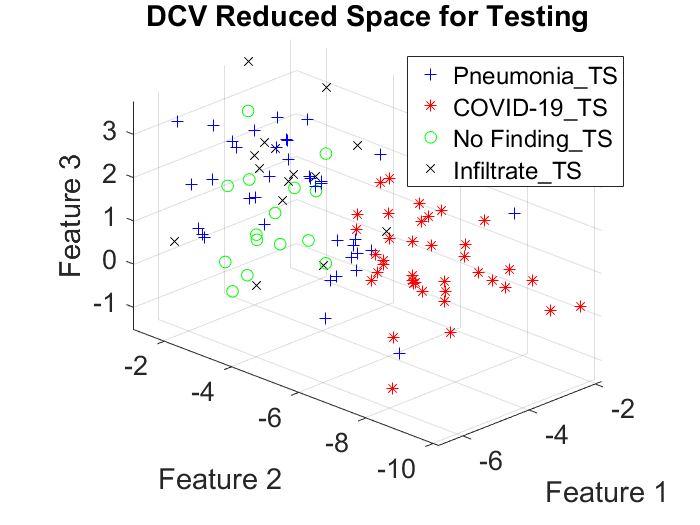}\\

\end{tabular}
     \caption{Point distribution in the reduced
     space given by the 3 principal components of each method.
     Distribution for training set in left side plots and
     for testing in right side plots.}
     \label{fig:RedDim}
\end{figure*}

Figure \ref{fig:RedDim} shows the point cloud in the space given
by the 3 principal components of each method for the training and
test sets for one of the folds. Test samples are labelled adding a
"TS" to the name of the class. Methods based on analysis of
principal components (PCA and KPCA) perform poorly already in
discriminating the training samples. This is expected given that
they compute axis capturing the larger variability among samples
and these axis are not always the ones that best separate classes.
These axis are usual part of the null space of the covariance
matrix \cite{Cevikalp05}. From the two discriminant methods (LDA
and DCV), DCV is the one achieving largest separability in
training samples thanks to a high fraction of variance that
controls the separability in the pseudo-null space. Given that DCV
class separability is preserved acceptably in the testing samples,
we have selected this method as the most suitable for
dimensionality reduction.

Average scores and their 95\% CIs for precision and recall for
each cell size are shown in Table \ref{Tab:CellSze}. The highest
precision is achieved by $CellSze=16$, with a $CI=[0.7586902,
0.9105691]$. This size achieves also the second highest recall
with $CI=[0.7999563, 0.9619484]$, which is very close to the
highest one achieved by $CellSze=4$. The p values for the pair
wise comparison of cell sizes and confidence intervals for the
difference for each score are shown in Table
\ref{Tab:CellSzeComp}. Significant effects are shown in bold face.
The precision achieved by  $CellSze=16$ is significantly better
than the one achieved by  $CellSze=4$ and recall is comparable.
Although it is not significant, $CellSze=16$ intervals comparing
its precision to the remaining cell sizes have a negative bias
and, thus, $CellSze=16$ tends to have a higher precision. This is
also the case for recall intervals. Therefore, we conclude that
$CellSze=16$ is the best suited size for HoG descriptor in images
of size $400 \times 400$.

\begin{table}[!h]
\centering
      \caption{Effect of HoG cell size in COVID-19 Detection: average scores}\label{Tab:CellSze}
      \begin{tabular}{r cc cc}
              $CellSze$ & \multicolumn{2}{c}{Precision}   &  \multicolumn{2}{c}{Recall} \\
             \hline & \multicolumn{1}{|c}{Average}  & \multicolumn{1}{c|}{95\% CI} &  Average  & 95\% CI   \\
             \cline{2-5} \multicolumn{1}{r|}{$CellSze=4$} & 0.7651227 & \multicolumn{1}{r|}{[0.6891832, 0.8410621]}
              & 0.8904762
            &[0.8094802, 0.9714722]
            \\   \multicolumn{1}{r|}{$CellSze=8$} & 0.8031650 & \multicolumn{1}{r|}{[0.7272255, 0.8791045]} & 0.8476190 & [0.7666230, 0.9286151] \\
            \multicolumn{1}{r|}{$CellSze=16$} & 0.8346296 & \multicolumn{1}{r|}{[0.7586902, 0.9105691]} &  0.8809524 & [0.7999563, 0.9619484] \\
            \multicolumn{1}{r|}{$CellSze=32$} &  0.7910462 & \multicolumn{1}{r|}{ [0.7151067, 0.8669857]} & 0.8317460 & [0.7507500, 0.9127421] \\
            \hline
      \end{tabular}
\end{table}

\begin{table}[!h]
\centering
      \caption{Effect of HoG cell size in COVID-19 Detection: comparison across sizes.
      Significant effects are shown in bold face}\label{Tab:CellSzeComp}
      \begin{tabular}{r cc cc}
              $CellSze$ & \multicolumn{2}{c}{Precision}   &  \multicolumn{2}{c}{Recall} \\
             \hline & \multicolumn{1}{|c}{p value}  & \multicolumn{1}{c|}{95\% CI} &  p value  & 95\% CI   \\
             \cline{2-5} \multicolumn{1}{r|}{$4-8$} & 0.2473 & \multicolumn{1}{r|}{[-0.12477274, 0.04868808]}
              & 0.1817            & [-0.04156023, 0.12727451]
            \\   \multicolumn{1}{r|}{$4-16$} & {\bf 0.0379} & \multicolumn{1}{r|}{[-0.15623738, 0.01722343]} & 0.7643  &
            [-0.07489356, 0.09394118]\\
            \multicolumn{1}{r|}{$4-32$} & 0.4285 & \multicolumn{1}{r|}{[-0.11265393, 0.06080689]} & 0.0698 &
            [-0.02568721 , 0.14314753]
            \\
            \multicolumn{1}{r|}{$8-16$} &  0.3374 & \multicolumn{1}{r|}{ [ -0.11819506, 0.05526576]} & 0.2969 &
            [-0.11775070, 0.05108404] \\
\multicolumn{1}{r|}{$8-32$} &   0.7105 & \multicolumn{1}{r|}{
[-0.07461160, 0.09884922]} &  0.6176 & [-0.06854435, 0.10029038]
\\
\multicolumn{1}{r|}{$32-16$} &  0.1861 & \multicolumn{1}{r|}{
[-0.13031386, 0.04314696 ]} & 0.1264 &
[-0.13362372,  0.03521102]\\

            \hline
      \end{tabular}
\end{table}

Table \ref{Tab:CellSzeComp} reports the quality scores to be
compared with the numbers in Table \ref{tb: experimental design}.
The proposed method achieves average scores above 90\% for all
configurations of class distribution which are competitive, even
better in some cases, to the deep learning approaches reported in
Table \ref{tb: experimental design}. Comparing to methods
considering only the Covid/Normal groups, the proposed method is a
bit better than \cite{narin2020automatic} in terms of recall, with
a 4\% drop in precision, and much better than
\cite{hemdan2020covidx} in terms of precision. Although the target
it is COVID-19 detection and, thus, recall should be high, in a
screening problem precision should also be high enough to avoid
unnecessary tests to confirm COVID-19. In comparison to methods
considering Covid/Pneumonia groups, ours is better than
\cite{zhang2020covid} in specificity with only 2\% in recall and
better than \cite{castiglioni2020artificial} in all scores. Since
scores are computed for a two class problem (COVID-19 versus non
COVID-19), being our method better in specificity implies that it
also has a higher precision. Finally, comparing to methods trained
to discriminate Covid/Pneumonia/Normal groups, our method is
comparable to \cite{apostolopoulos2020covid} in terms of recall
with better precision and much better in precision than
\cite{wang2020covid} with only 4\% drop in recall.

\begin{table}[!h]
\centering
      \caption{Comparison to SoA methods in COVID-19 Detection.}\label{Tab:CellSzeComp}
      \begin{tabular}{r cc cc}
 Classes & Accuracy    &  Sensitivity(Recall) & Specificity & Precision\\
            \hline
\multicolumn{1}{r|}{COVID-19/Normal} & 96 &98 & 93 & 96 \\
\multicolumn{1}{r|}{COVID-19/Pneumonia} & 93 & 94 & 92& 95\\
\multicolumn{1}{r|}{COVID-19/Pneumonia/Normal} & 96 &98 &93 & 96\\
      \end{tabular}
\end{table}

Finally, regarding COVID-19 early detection the analysis of
variance did not detect any significant differences across the 3
groups with a p-value equal to 0.8284 and average detection rates
equal to 89\%, 93\% and 94\% for, respectively, early, mid and
late COVID-19 cases.

\section{Discussion and Conclusions}\label{sc:discussion}

In this paper we have presented a method for COVID-19 detection in
X-ray based on HoG and reduction of dimensionality. Model
parameters include, both, reduction method and HoG cell size and
were tuned using statistical analysis of the results obtained for
the classification of COVID-19, non-COVID-19 pneumonia,
non-COVID-19 infiltration and normal cases. The most suitable
configuration was DCV dimensionality reduction and a cell size
equal to $16 \times 16$ pixels. This model was assessed in 2
aspects: comparison to state-of-art deep learning methods and
capability for early detection of COVID-19. The results of our
experiments raise some interesting points.

The proposed classic approach is comparable (even better in many
cases) to deep learning methods. The accuracy for discriminating
among COVID-19, non-COVID-19 pneumonia and normal cases is above
95\% with a recall of 98\% and a precision of 96\%. None of the
existing methods included pulmonary infiltration as class, which
according to our experiments and clinical evidences might be
introducing a positive bias in quality numbers.

Comparing the above results to the accuracy for discriminating
between COVID-19 and non-COVID-19 infiltration carried out for
model parameters tuning, we have that non-COVID-19 infiltration is
the group that most confuses with COVID-19. From a clinical point
of view this is expected due to the fact that most non-COVID-19
pneumonia are bacterial and bacterial pneumonia are radiologically
very different to a viral one like COVID-19. Viral pneumonia looks
like bilateral infiltrations in X-ray. The problem with this is
that infiltrations occur in other non-COVID-19 pathologies (acute
lung edema, other non-COVID-19 viruses, respiratory distress,
etc). Although according to our experiments the number of
infiltrations wrongly classified as COVID-19 represent less that
20\% of the cases detected as COVID-19 by the system, we consider
that theses cases should be further filtered using other clinical
variables like signs of heart failure (by medical history,
physical examination, or analytic).

Regarding early COVID-19 screening using X-ray, our experiments
indicate that COVID-19 detection rates is similar at early, mid
and late stages of the pathology. However this conclusion should
be carefully confirmed using data properly recorded, since the
stage of COVID-19 was determined by the number of days past
between the start of symptoms or hospitalization and acquisition
of each image. Without further radiological description and
clinical data this does not guarantee that the first image was
acquired at an early stage.

Still, results are very encouraging and, in our opinion, validate
the feasibility of early COVID-19 screening using X-ray and,
possibly, other clinical variables. The immediate step to fully
confirm the actual clinical benefits of this screening is to test
our method in retrospective cases collected from Catalan primary
care centers.

\section*{Acknowledgments}
The research leading to these results has received funding from
the European Union Horizon 2020 research and innovation programme
under the Marie SkAodowska-Curie grant agreement No 712949
(TECNIOspring PLUS) and from the Agency for Business
Competitiveness of the Government of Catalonia.

This work was supported by Spanish projects RTI2018-095209-B-C21,
FIS-G64384969, Generalitat de Catalunya, 2017-SGR-1624 and
CERCA-Programme. Debora Gil is supported by Serra Hunter Fellow.
The Titan X Pascal used for this research was donated by the
NVIDIA Corporation.

"Dedicat a la mama (DGil)"

\bibliographystyle{unsrtnat}
\bibliography{XRayScreening_Arxiv}

\end{document}